# Superconducting source and sensor of single bubbles to study nucleate boiling of liquid helium

A.G. Sivakov[1], O.G. Turutanov[1,2*], S.A. Kruhlov[1], A.V. Krevsun[1], A.S. Pokhila[1], A.E. Kolinko[1], S.I. Bondarenko[1], M. Grajcar[2]

[1]B. Verkin Institute for Low Temperature Physics and Engineering of the National Academy of Sciences of Ukraine, 47 Nauky Ave., 61103 Kharkiv, Ukraine

[2]Comenius University Bratislava, Mlynská dolina F1, 842 48 Bratislava, Slovak Republic

*e-mail: turutanov@ilt.kharkov.ua

**Abstract**

Joule heat generated by resistive elements of cryogenic micro- and nanodevices often originates boiling of the cooling cryogenic liquids (helium, nitrogen). The article proposes an experimental method to explore the dynamics of the formation and development of a single vapor bubble in cryogenic liquid by sensing the temperature change of a superconducting thin-film microbridge being in the resistive state with single phase-slip center or line. It serves both the source of heat for generating single bubbles and the surface temperature sensor due to its temperature-dependant excess current. The average bubble detachment rate and the average single bubble volume were experimentally determined for nucleate helium boiling. The obtained values are in good agreement with the data of other authors found in literature.

## 1. Introduction

A researcher studying the resistive state of superconducting films inevitably faces the problem of Joule heating of a sample that can considerably influence the current-induced destruction of superconductivity. It also narrows the measurement range, since the superconducting current depends on temperature and vanishes above the critical temperature. The same problem exists in applications of superconducting micro- and nanoelectronics containing resistive elements. Therefore, the conditions for good heat removal play an important role. They can be improved by using substrates with high thermal conductivity (quartz, sapphire), with good acoustic matching between the film and the substrate, with a small cross section of the structure, providing two-dimensional or three-dimensional phonon propagation, etc. When the sample is placed in a cryogenic liquid (liquefied nitrogen, helium), other nonlinear channels of heat removal appear provided by thermal conductivity of the liquid, convection, and various boiling regimes.

The technical significance of the boiling in various liquids, primarily water, is extremely high, and always attracted the attention of engineers and researchers. A big leap in understanding of the heat removal mechanism in boiling water occurred almost 90 years ago. Then, trying to find a way to speed up the heating of the boilers of Japanese warships, Shiro Nukiyama experimentally explored various regimes of water boiling using a platinum wire heater. He has determined the points of boiling crises and corresponding critical heat fluxes (CHFs) from a heated surface [1], and depicted currently well-known "boiling curve". It is an N-shaped dependence of the heater specific power $q$ on the temperature difference between the surface and the liquid layer $\Delta T$ (Fig. 1).

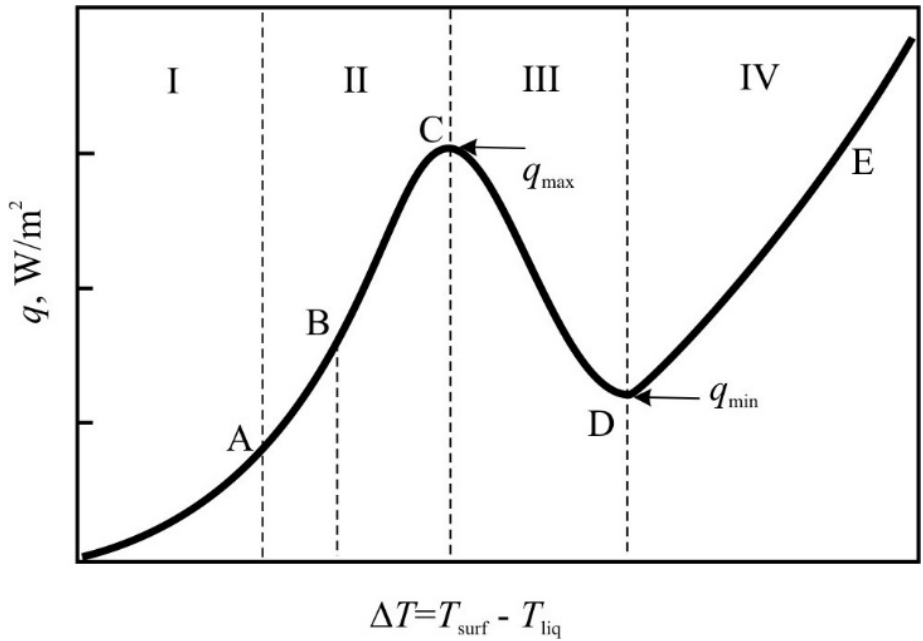

Fig. 1. Generalized boiling curve for liquids, which is heat flux from the heated surface into the liquid $q$ vs. temperature difference between the surface and the adjacent liquid layer $\Delta T$. Boiling regimes: I—convective boiling, II—nucleate boiling, III—transitional regime, IV—film boiling. Section A-B corresponds to boiling by individual bubbles, C and D are boiling crisis points.

The boiling curve contains sections of convective heat removal, nucleate boiling, transition boiling, and film boiling regimes. This is also true for cryogenic liquids. The nucleate boiling regime provides the best heat removal from the heated surface with minimal temperature difference between the surface and the liquid. The specific values of the critical heat fluxes and the superheat $\Delta T$ depend on many factors, including the surface material. For example, the experiments with liquid helium have shown that the temperature difference corresponding to the first boiling crisis is $\Delta T_{cr1} = 0.17$ K for a copper, and $\Delta T_{cr1} = 0.32 - 0.38$ K for bronze, nickel and stainless steel surfaces [2].

The investigations of boiling of cryogenic liquids are important for the development of low-temperature electrical and electronic engineering, as well as space applications. Cryogenic liquids have some specific features like almost zero wetting angle. Due to good wetting of the surfaces, the bubble formation mechanism in this case differs from that of common liquids [2, 3]. Boiling in cryogenic liquids has been studied theoretically and experimentally over the past few decades [2–8]. Nevertheless, appearance of recent works, e.g. [9–11], indicates that the problem is still open and the research is incomprehensive. In the case of liquid helium there are significant experimental and theoretical difficulties, as compared to conventional liquids.

With the development of micro- and nanotechnologies, the size of film structures that are in resistive state and may serve as nuclei for the vapor bubble formation are getting smaller than the single bubble size. In applications and research studies, this leads to unacceptably large temperature fluctuations of the microstructure. For example, a special study [12] was devoted to oscillations of differential resistance of current-driven microscopic point contacts caused by helium-II boiling. At the same time, the effect luckily provides a research tool to study experimentally the dynamics of the temperature fluctuations caused by the evolution a single vapor bubble. The bubbles start to appear after the first CHF is reached, manifesting convective to nucleate boiling transition, unlike common approach of "pool boiling" in which formation of multiple bubbles in bulk liquid helium volume is considered.

To enable studies of single-bubble helium boiling in rather large temperature range, we propose to use a short narrow thin film microbridge for both generation and detection of single helium bubbles by monitoring its temperature change.

## 2. Thermal sensitivity of superconductors

In general, superconductors are well suited for the purposes of thermometry, since many of their parameters (critical current, resistance in the vicinity of the superconducting transition) dramatically depend on temperature. Latter feature is used, in e.g. superconducting bolometers with transition-edge sensors (TES) [13], level gauges [14], and underlies the technique of laser scanning microscopy of superconductors [15]. But in these cases, a narrow temperature interval is exploited near the superconducting transition, although there are other applications based on the temperature dependence of the critical current. A larger temperature range is needed to study the temperature dynamics of nucleate boiling.

The larger temperature range can be achieved by using some properties of the resistive state of thin-film superconductors with centers [16] and lines [17] of the superconductor order parameter phase slip (PSCs and PSLs, respectively). Namely, the localized regions with nonzero resistance and the so called temperature-dependent "excess current" in the current-voltage characteristics (or I-V curves, IVCs) are their specific features. Starting from the first works [16,18] devoted to the PSCs in narrow superconducting channels and "hot spots", the authors emphasized noticeable improvement in heat removal from a short narrow thin film bridge ($0.5 \times 0.5$ μm) due to nucleate boiling, which was seen in the IVCs as a change of their shape, in agreement with the simple theoretical model of heating [18]. This gives us the basis for the temperature "microsensing" during helium nucleate boiling.

Note that, up to the date, the boiling of cryogenic liquids was investigated with large-area heated surfaces generating multiple bubbles, and so averaged heat fluxes were measured as a result.

## 3. PSC/PSL as a temperature sensor

According to [16], when the current exceeds the depairing critical current, a local resistive region abruptly emerges in a narrow superconducting thin-film strip, which is a quasi-one-dimensional superconducting channel from the theoretical [19] point of view. Periodic oscillation of the order parameter with Josephson frequency occurs in the center of this region generating excess nonequilibrium quasiparticles. These quasiparticles diffuse in both directions along the superconducting channel to a certain depth, which is determined by the Fermi velocity $v_F$, mean free path $l$ and inelastic relaxation time $\tau_{ph}$. They drag the electric field with themselves, and finally come to equilibrium with the superconducting condensate. The penetration depth of the electric field $l_E$ in traditional low-temperature superconductors has an order of one to tens microns (4-6 microns for tin). In later microscopic theories [19], $l_E$ slightly depends on temperature $T$, $\sim (1 - T / T_c)^{-1/4}$, where $T_c$ is the critical temperature. This $2l_E$-long region of charge imbalance generates Joule heat. At the same time, anharmonic oscillations of the order parameter in the middle of the PSC cause the alternate flow of superconducting and normal currents through the channel. When averaged over time, this resulted in the IVC with the linear part with differential resistance $dV / dI = R_N \frac{2l_E}{L}$, where $L$ is the total length of the channel and $R_N$ its resistance in normal state. Each successive PSC entered the channel adds $R_N \frac{2l_E}{L}$ to the differential resistance up to the moment the channel is fully filled with PSCs, and then the linear part in the IVC parallel to normal resistance line reaches differential resistance $dV / dI = R_N$. If continued to intercept with the current axis, the line cuts off a segment $(0.5 - 0.7) I_c$ on the axis [16]. This value is called excess current $I_{exc}$ and depends on temperature. Theoretically [19], in a long quasi-one-dimensional superconducting channel with width and thickness $w, d < \xi(T)$, the excess current should exist up to so called second critical current density $j_{c2}$ (the current of instability of normal state to the formation of superconducting nuclei) exceeding critical current density $j_c$ by an order of magnitude, but this region is actually limited by thermal effects [20]. As the current increases, new PSCs successively enter the channel, and the total current change occurs due to the normal current component only, while the average superconducting current remains unchanged. Thus, periodically arranged local resistive domains of length $2l_E$ appear in the long superconducting channel. They dissipate heat, the excess current in the IVC of the channel being a function of the temperature of these domains. Although PSCs are non-stationary non-equilibrium domains with complicated structure, we can consider them for our sake just as "almost normal" heat-dissipating thin film parts which are also sensitive to the temperature. Fig. 2 shows an initial section of the schematic I–V characteristic of a long superconducting channel which can contain up to four PSCs. The IVC asymptotically tends to the normal resistance line at higher currents (not shown). The inset shows resistive domains of $2l_E$ in length distributed in the long channel when two and four PSCs enter the channel, correspondingly. As the channel length increases, the number of voltage steps increases, too, and the steps become indistinguishable in the end. In the case of relatively wide film strips ($w > \xi(T)$ and even $w, d > \lambda(T)$, where $\lambda(T)$ is the magnetic field penetration depth), phase slip lines (PSLs) [17] rather than PSCs appear in the strip demonstrating similar IVC behavior.

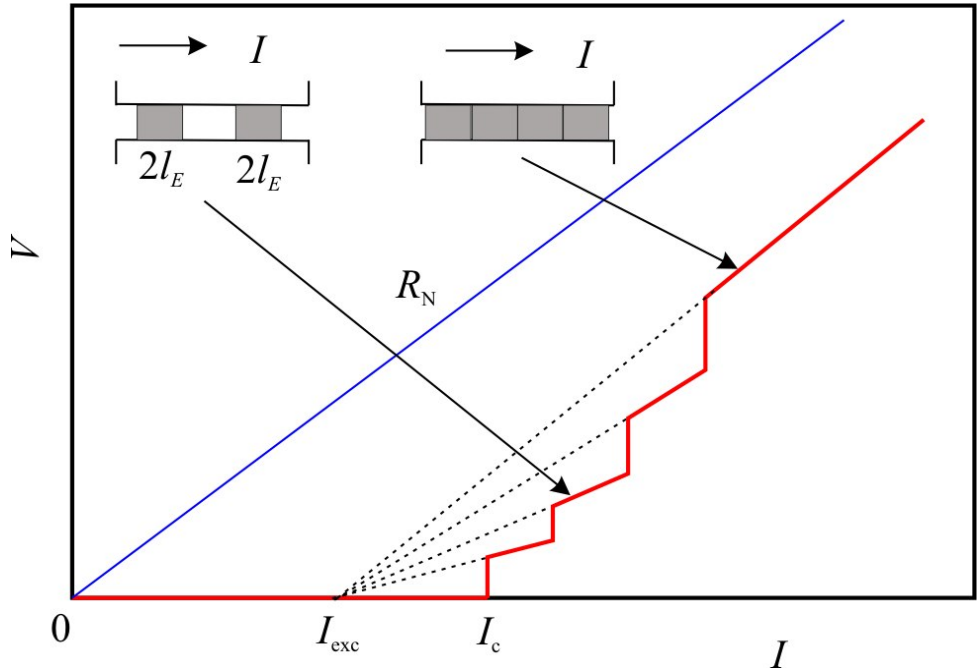

Fig. 2. Schematic I–V curve for a narrow superconducting channel with four PSCs. The inset shows resistive domains of $2l_E$ in length distributed in the channel for two and four PSCc entered the channel.

Thus, the IVC at high current can be described by a simple asymptotic expression $I = V/R_N + \alpha I_c$, where $\alpha = 0.5 - 0.7$ and $I_{exc} = \alpha I_c$, if not taking into account thermal effects [20]. With the length of a superconducting microbridge $L \leq 2l_E$, only one PSC (or PSL at $w \gg \xi(T)$) can be housed in it. Note that its size does not exceed the critical diameter of a single vapor bubble in liquid helium [4]. The IVCs of a tin thin film microbridge with length approximately equal to the size of single PSC (PSL) are shown in Fig. 3(a).

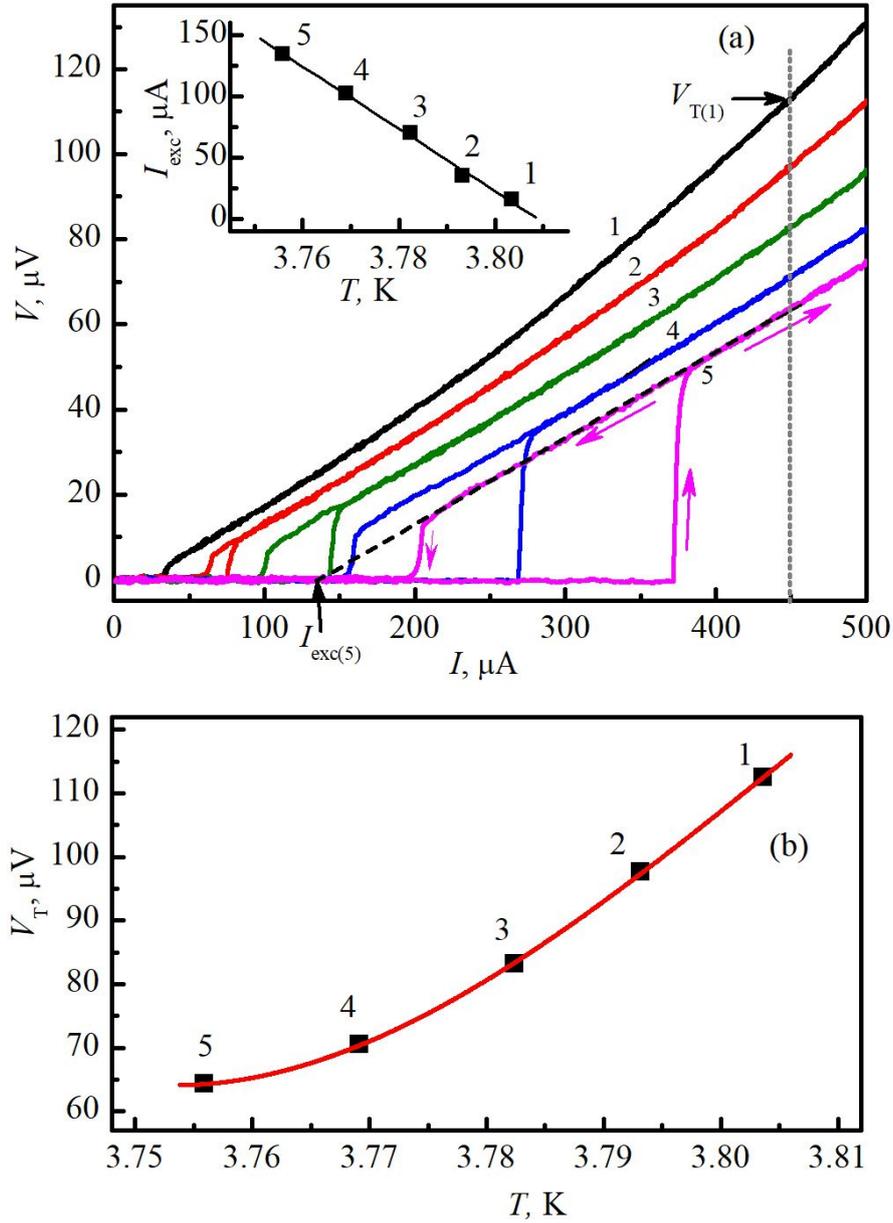

Fig. 3. Short superconducting tin thin film microbridge $5\times1\times0.2$ μm in size deposited on crystal quartz substrate as a temperature sensor: (a) Set of IVCs for various temperatures: 1—3.804 K, 2—3.793 K, 3—3.782 K, 4—3.769 K, 5—3.756 K (critical temperature $T_c = 3.81$ K). The inset shows excess current vs. temperature, the numbered points correspond to the curves numbers at the main panel. (b) Temperature dependence of the dc voltage across the bridge $V_T$ taken at a fixed current of 450 μA (shown by dashed line in panel (a)). The points enumeration is the same as in (a).

One can see that the power generated by the sample can be varied over a wide range, keeping the superconducting state with an excess current. Its temperature dependence $I_{exc}(T)$ is shown in the inset. The change in the sample temperature under given conditions of heat sink can be easily tracked by dc voltage $V_T(T)$ taken at a fixed current (Fig. 3(b). Since this fixed current value is chosen rather high (greater than the largest $I_c(T)$ for shown IVCs), the nonlinearity of $V_T(T)$ is explained by overheating effect. Knowing $I_{exc}(T)$ and calibrating the corresponding $V_T(T)$, we can measure the sample heating over the helium bath during the single vapor bubble nucleation and detachment.

In the above case, a thin-film sample was deposited on a quartz substrate, which had good thermal conductivity and good acoustic matching with the film, so was able to remove significant part of the heat. To make heat removal determined mainly by the nucleate boiling mechanism, the thermal film-substrate coupling must be weakened, and the substrate itself must have low thermal conductivity.

## 4. Experimental results and discussion

To create a detector for measuring the temporal and thermal characteristics of single-bubble evolution during nucleate boiling of liquid helium at atmospheric pressure, we have taken a superconducting thin film bridge with size of about $10\times10\times0.15$ μm made of InSn alloy (50% + 50%) with critical temperature ~6 K, i.e. above the helium boiling point. The film was deposited on a glass-ceramic substrate of $5\times10\times0.5$ mm in size. The ceramic substrate was chosen in order to significantly reduce the heat flux into the substrate compared to the heat flux into liquid helium. In our experiments, we just demonstrated the principle of constructing a superconducting source and detector of single bubbles, so we have not measured the heat flux into the substrate and into the electrical leads. All further measurements were made in liquid helium at atmospheric pressure, i.e. at temperature ~4.2 K. The scheme of the sample is shown in the inset in Fig.4. Several samples were made which were slightly different in size and thermal coupling.

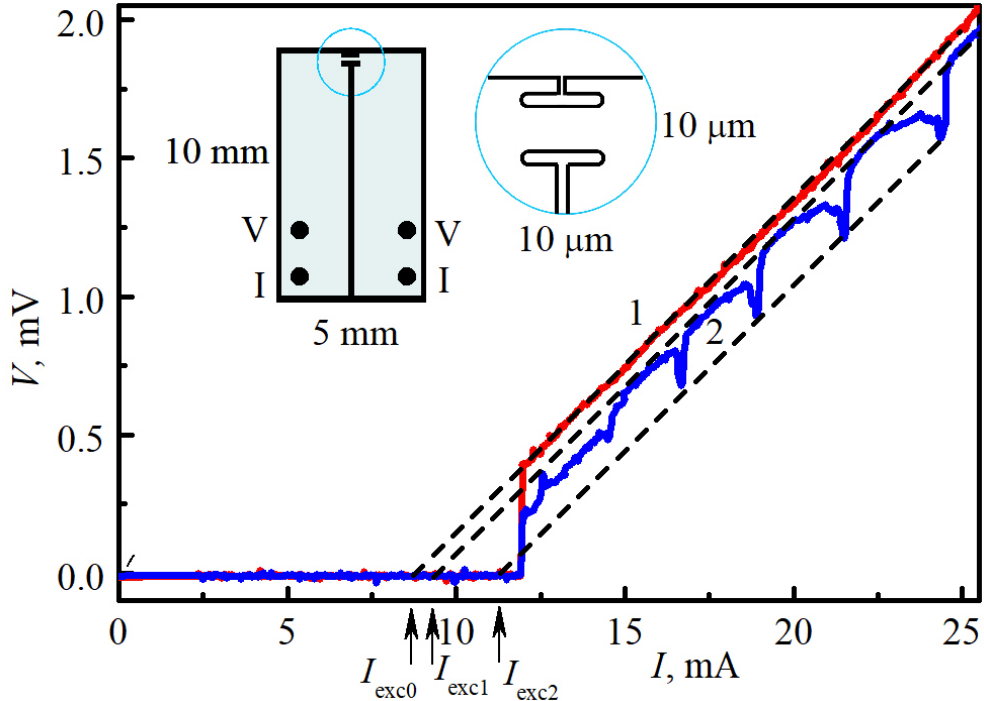

Fig. 4. Sensing liquid helium level: IVCs of a thin-film bridge taken in gaseous (curve 1) and liquid (curve 2) helium. Excess currents $I_{exc}$ are shown by arrows on the current axis (explained in the text). The orientation of the substrate is vertical. Deposited thin film is InSn (50%+50%) alloy 150 nm thick on a glass-ceramic substrate. The inset shows a diagram of the sample indicating the location of the bubbles detector.

Note that the critical current of a thin-film bridge depends linearly on the cross section of the sample, and the normal resistance is inversely proportional to the cross section, while the generated power is proportional to the current squared. Thus, by varying the size of the bridge, it is possible to control the specific power per unit area to provide the nucleate boiling conditions keeping in mind that the bridge length is limited from above by $2l_E$ to ensure that a single PSC or PSL fit in the sample. Experimentally, this is easily verified by the I–V characteristic, which should contain only one voltage jump and one linear section with excess current.

Fig. 4 displays the IVCs of the microbridge recorded in gaseous helium (curve 1) and in liquid helium (curve 2). Nucleate helium boiling causes temperature fluctuations that result in voltage oscillations in resistive section of IVC due to changes in excess current $I_{exc}(T)$ during slow IVC recording. In contrary, there is no voltage oscillation in the IVCs when the sample is exposed to the gas.

The excess current is determined by the segment cut off at the critical current by extended linear section of the IVC at a given temperature. The values of excess currents at the extreme points of temperature fluctuations are indicated by arrows. Note that critical currents for both curves are equal because there is no dissipation in the superconducting state.

Thus, the microbridge detector sensitively distinguishes the change in heat removal conditions between gaseous and liquid phases working as a liquid helium level meter.

IVCs taken in a wide range of currents in a linear section with excess current at different recording rates (Fig. 5) demonstrate two facts: the oscillation frequency (a) is practically independent of the current value and (b) does not depend on the current sweeping rate during recording the IVC. The last fact is quite understandable since the IVC record time is 1 s, much greater than characteristic time of the bubble formation and detachment which, as we will see below, is of order of 0.01 s. So we can consider the IVCs as "adiabatic curve" in this sense.

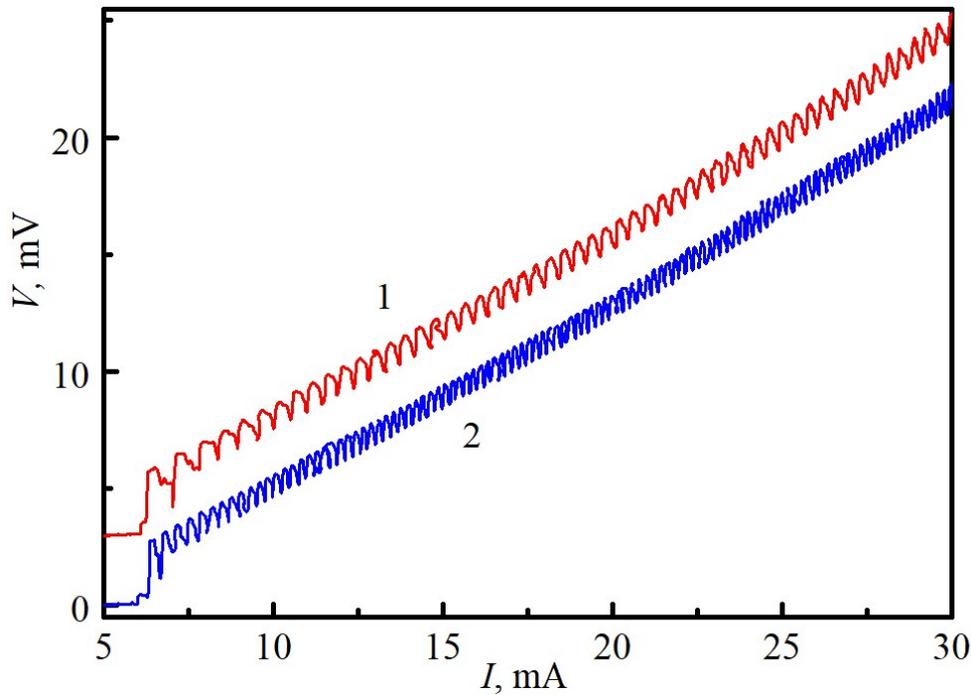

Fig. 5. Resistive parts of the IVCs of the microbridge recorded in a wide current range with different rates: 2.5 s (curve 1) and 5 s (curve 2). Curve 1 is shifted up by 3 mV for clarity. The orientation of the substrate is vertical.

The fact that the oscillation frequency is almost independent of the current is also confirmed by the voltage across the bridge in time domain ("oscilloscope mode") obtained at three different currents (Fig. 6(a)). Note that this sample had the worst thermal coupling with the substrate, therefore curves 2 and 3 correspond to a significant overheating of the film. Nevertheless, the main oscillation frequency, determined from the Fourier spectra of these time series (Fig. 6(b)), is constant and equals 58 Hz for this case (marked with dashed vertical line). This value is in good agreement with the experimental data [4] on the detachment rate of vapor bubbles from a heated surface in liquid helium, where it was 135–204 Hz, especially taking into account differences in the experimental setup and statistical spread of the detachment rate in each experiment.

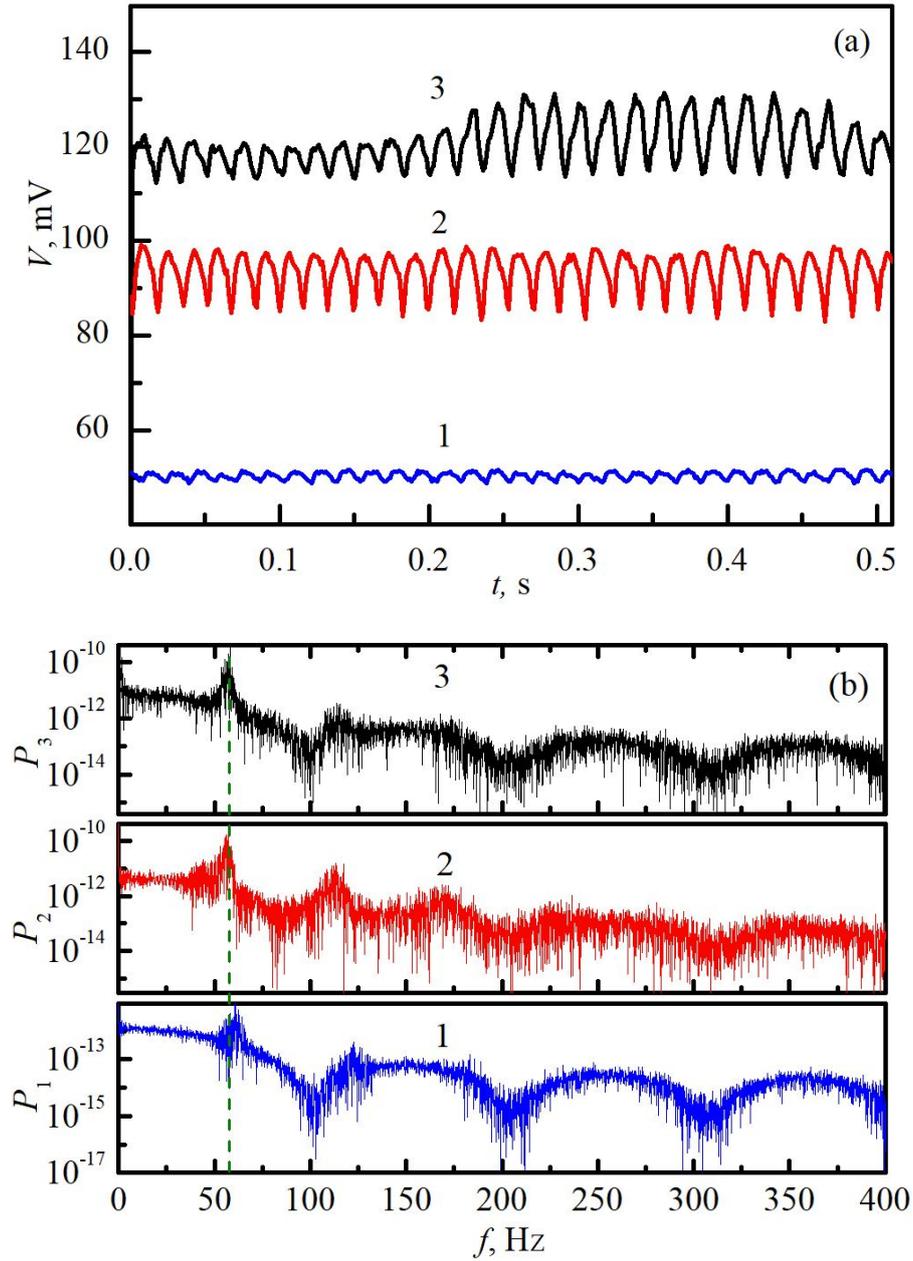

Fig. 6. Bubble detachment frequency: dc voltage across the microbridge vs. time at various fixed currents: 1 - 28.8 mA, 2 - 34.95 mA, 3 - 40 mA (a) and the corresponding Fourier spectra. The fundamental bubble detachment frequency is indicated by the vertical dash line (b).

Thus, each oscillation can be unambiguously associated with the act of formation and detachment of a single bubble. The shape of an oscillation is complex and can vary since the heat removal during the growth and detachment of a bubble changes nonlinearly. The constancy of the bubble detachment rate, regardless of the power, means that the volume of each detached bubble is proportional to the power. This agrees with the data on helium boiling [2] which show that the portion of larger bubbles leaving the heating surface increases as the heat flux rises, while the total number of bubbles remains (in visual observation) almost unchanged. The increased amplitude of oscillations can be ascribed to the deterioration of heat removal due to larger bubble size in one bubble growth cycle.

Expressions like $fD_0^n$, where $f$ is the bubble detachment rate and $D_0$ is the bubble detachment diameter, are used by some researchers for the theoretical analysis of heat transfer in bubble boiling regime of liquids [2]. The value of $n$, accepted by various authors, is usually in the range of 0.5–3.

The above scenario of bubble formation with the dissipated power rise in the microbridge is in good agreement with the assumption that $fD_0^3 = \text{const}$. The value $fD_0^3$ can serve as a characteristic that determines the volumetric vapor velocity generated by the center of vaporization, the microbridge in this case.

Fig. 7 shows an enlarged fragment of one of the voltage oscillograms and a schematic diagram of single bubble evolution when the bubble goes through several stages.

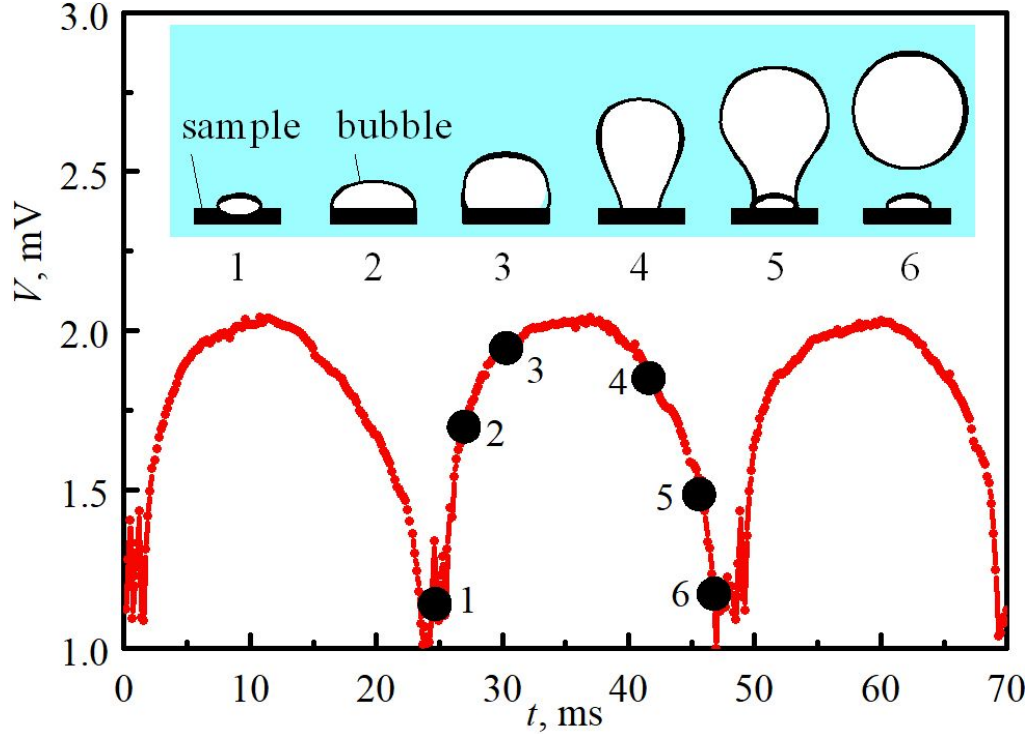

Fig. 7. Bubble evolution: dc voltage across the microbridge vs. time for several oscillations at a fixed current of 7 mA. Numbered dots mark different stages of vapor bubble development on the surface of the sensor. The inset shows a schematic diagram of the vapor bubble evolution.

Upon reaching a certain specific power generated on the sample surface, which corresponds to point A in Fig. 1, the convective heat removal is replaced by the formation of a vapor bubble (point 1 in Fig. 7 and inset). The thermal conductivity of the gas is lower than that of the liquid, so the heat removal from the sample deteriorates, and its temperature increases. This leads to a decrease in excess current and voltage rise. Further, the bubble grows and quickly occupies the entire area of the sample (the wetting angle for cryogenic liquids is close to zero), which leads to a further temperature rise (point 2). The volume of the bubble increases, the heat removal falls down even more; the temperature and voltage become higher (point 3). With larger diameter of the bubble, the buoyancy force increases, and the bubble is elongated, forming a stem. The area wetted by the liquid increases, and the temperature begins to fall (points 4, 5). When the bubble detaches from the surface, the liquid cover the sample again, and the temperature drops down, so the excess current increases and hence the voltage goes down (point 6). Then the process repeats. This behavior generally corresponds to the model of development of a single bubble [2].

The critical size and the bubble detachment rate are determined by the balance of forces that bind the bubble to the surface and the buoyancy force, and depend on many parameters, the surface cleanliness and wettability, its curvature, the characteristics of the coolant, etc. Among these parameters, the orientation of the surface where the bubbles nucleate since cooled technical elements in applications may have any orientation.

In Fig. 8 one can see oscillograms of the output signal (detector voltage) for two cases of the substrate orientation, vertical (curve 1) and horizontal, the detector face down, i.e. above the bubble (curve 2). These cases are extreme from the point of view of the heat removal.

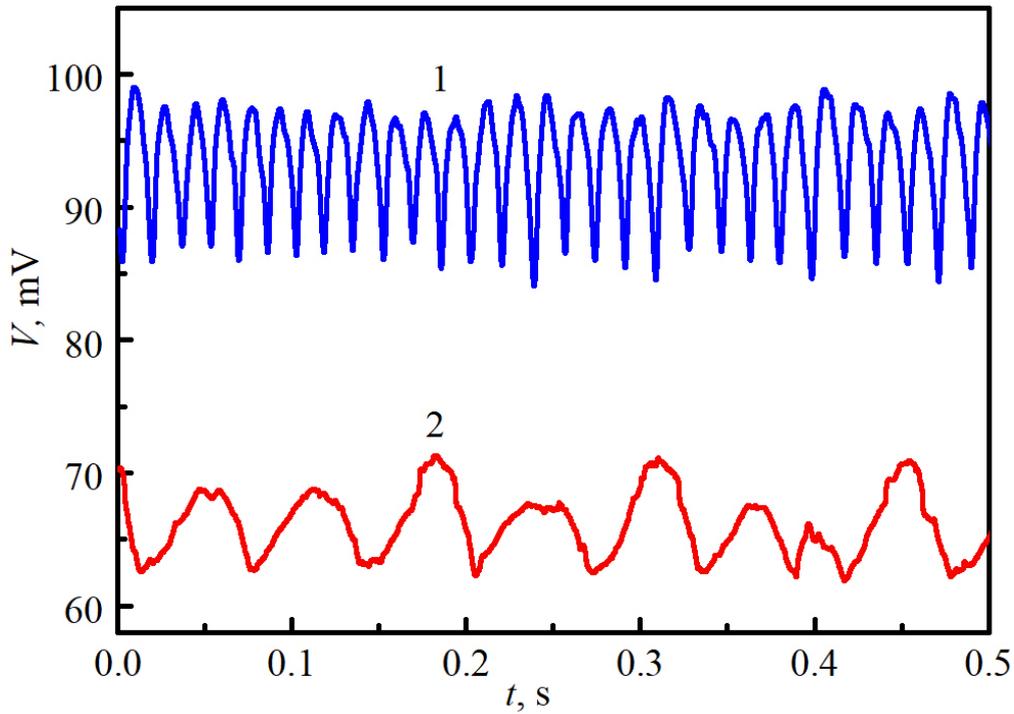

Fig. 8. Different substrate orientation: dc voltage across the bridge vs. time for vertical (curve 1, current 34.95 mA) and horizontal (detector face down) (curve 2, current 32.9 mA) substrate orientation.

To be more convincing in demonstrating the difference in heat removal, we placed the substrate horizontally, with the detector face downward. In this case, the bubble under the heated surface may not rise for a long time, increasing in size, although this effect is not fully realized since this surface is "half-open", the sensor is located on the edge of the substrate (see inset in Fig. 4). Despite the similar values of the power generated in the detector for these two cases, the gravity dramatically changes dynamics of the bubble formation and detachment. Apparently, with a vertical arrangement, convective flows and asymmetry of forces during the separation of the bubble stem greatly facilitate this process. As a result, the bubble detachment volume becomes smaller, while the detachment rate increases correspondingly to maintain the power balance. The different amplitudes of the oscillations on the curves are due to the nonlinearity of the voltage across the detector vs. power because of weak film-substrate thermal coupling. To measure exact temperature difference, a calibration procedure must be carried out.

It should be emphasized once again that above results are obtained just to demonstrate the applicability of the described detector to the experimental studies of single-bubble boiling.

Theoretical calculations of the bubble detachment diameters are very complicated. The models should consider the balance of forces including the surface tension, the lifting force, the viscosity force, the forces due to the convection of the liquid near the bubble, the inertial forces arising in the liquid during the vapor bubble growth, etc. A complete analysis of all these factors is extremely complex, so a number of simplifications are used in deriving relationships for detachment diameters $D_0$ [2]. In this regard, new experimental data to verify these models are highly needed.

The ability of the described device to serve simultaneously as a source and a counter of single bubbles, and a level meter as well enables setting up an experiment to determine the average bubble size. Fig. 9(a) schematically shows successive steps of such experiment. A top-end-closed cylindrical cap with inner diameter of 5 mm, was placed in a liquid helium dewar. A detector was mounted inside this cap. The distance from the detector to the cap cover was 5 mm. The cap contained a thin capillary in its top with a valve connected to the space above the liquid helium bath. Before starting the experiment, the valve was opened for a while to fill the cap with liquid helium to the top, after which the valve was closed again. For simplicity, the capillary with the valve is not shown in the figure. Then the detector was driven with current, becoming a source for formation of single bubbles, which the detector recorded by itself. Floated and merged bubbles gradually displaced liquid helium from the space under the cap until its level dropped down to the sensor location (phases I–IV). With known volume under the cap and the calculated number of bubbles, the average bubble size can be calculated.

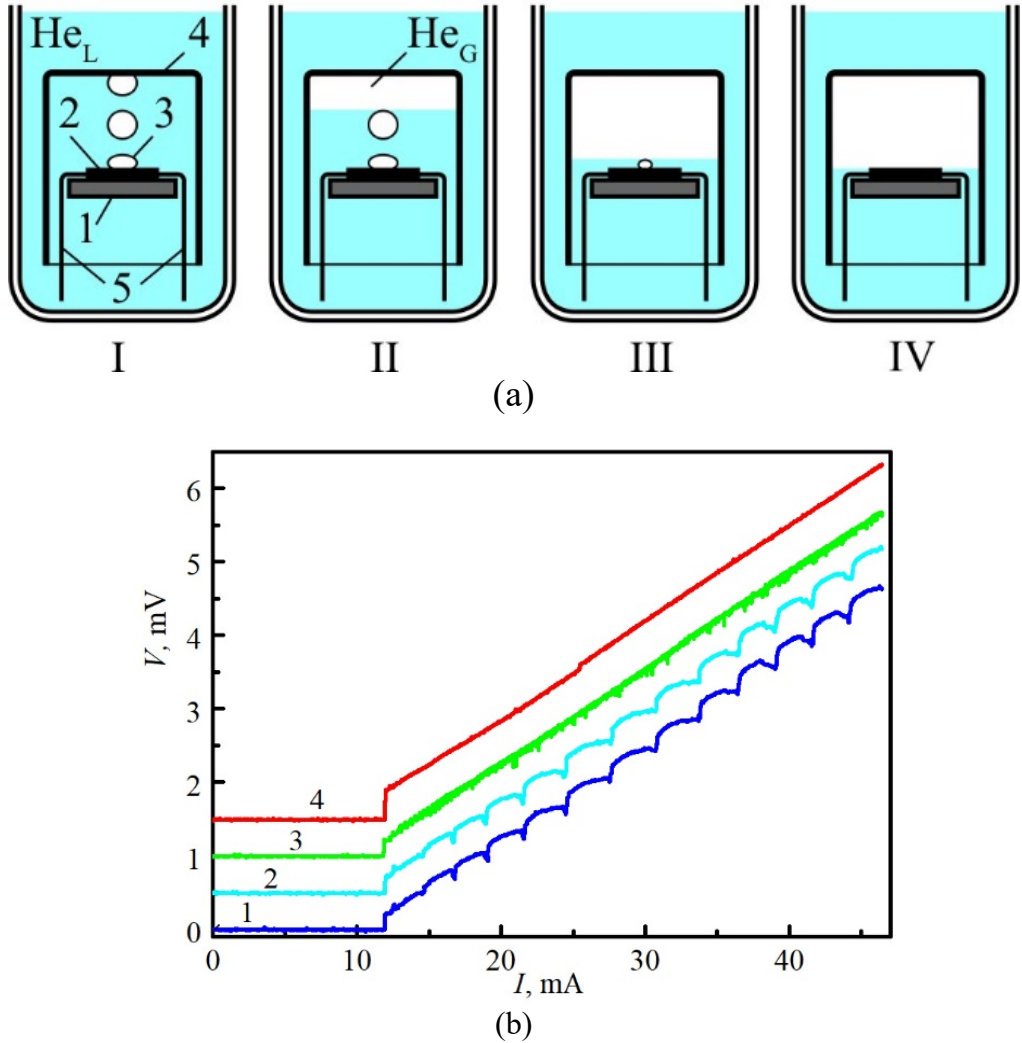

Fig. 9. Scheme of the experiment to determine the average volume of a single vapor bubble with different stages of boiling helium under the cap: I— boiling starts, II—the cap is partially filled with gas, III—liquid helium level dropped to the sample, IV—the sample is in the gas. In the first picture, the numbers indicate: 1 - substrate, 2 - superconducting thin-film bridge, 3 - helium bubble, 4 - cap, 5 - electrical leads from the sample (a). IVCs of the detector (curves 1–4) corresponding to boiling-out stages I–IV in panel (a) are shown (b).

In this experiment, we monitored the IVCs of the microbridge immersed in liquid helium under the cap until the bubble-induced oscillations disappear indicating the liquid helium level was below the sample. The IVC recording time was 1 s. Figure 9(b) shows the IVCs (curves 1–4) corresponding the stages I—IV of helium boiling shown in panel (a). One can see that full-fledged voltage oscillations turn into noise when the liquid helium level is close to the sample surface when boiling becomes irregular. The oscillations disappear when the liquid level reaches the detector surface. We see that the detector senses the moment when the liquid helium level reaches its surface that enables the bubble volume assessment. To repeat the experiment, the accumulated gas should be released through the capillary in the lid.

Fixing the current at 30 mA we counted 21923 oscillations in the oscillogram until the liquid helium level reaches the detector. With the volume under the cap of $\sim 100 \text{ mm}^3$, we get the average single bubble volume $\sim 0.0046 \text{ mm}^3$, and the average detachment diameter $D_0 \approx 0.17 \text{ mm}$. These values are in very good agreement with the experimental study of the growth dynamics of helium vapor bubbles at atmospheric pressure using high-speed filming [4], where the bubble detachment diameters were obtained $D_0 = 0.13 - 0.17 \text{ mm}$. The agreement is even more surprising if we take into account that this parameter has statistical character, and the scatter in the detachment diameters of vapor bubbles originated from a single vaporization center reaches 100% [2]. According to [2], the average size of the bubbles leaving the heated surface increases with the growth of the heat flux, while the total number of bubbles remains practically unchanged (which is also consistent with our measurements, see Figs. 5 and 6).

The above preliminary results show the practical applicability of the described technique for investigation of the single-bubble nucleate boiling in liquid helium. Additionally, such sensor can be used as a liquid helium level meter even in small volumes.

Further improvement of this method is possible in several directions. By using high-$T_c$ superconductors in this technique, one can explore the nucleate boiling regime in other cryogenic liquids with higher boiling point. A set of individual sensors or deposition of a matrix of thin-film sensors directly on the surface, possibly of a complex shape [11], gives a possibility of modeling and controlling multiple vaporization centers on a heated surface.

## 5. Conclusion

The paper considers an experimental method for studying heat removal from superconducting thin-film bridges of small (micron) sizes into liquid helium by single bubbles. Specific features of the resistive state of superconducting thin film bridges with formation of phase slip centers or phase slip lines open the way for creating a tool to study the single bubble evolution dynamics in nucleate boiling regime of cryogenic liquids. The local heating of PSC or PSL region initiates the nucleation, development, and detachment of a single bubble, so we have a "single-bubble source". At the same time, the excess current in a wide current range observed in IVCs depends on temperature and therefore is a good thermometric parameter. These properties underlie the proposed experimental method for studying the dynamics of single bubble evolution in cryogenic liquids. The measurements show voltage oscillations in IVCs of the superconducting microbridge immersed in liquid helium. They are unambiguously associated with the formation and evolution of single helium vapor bubbles. The shape of an oscillation reflects the temperature dynamics of the detector during formation, development, and detachment of a single bubble. The average detachment rate and average bubble diameter obtained experimentally are in very good agreement with the data of other authors taken by high-speed filming. Also, this detector can serve as a liquid level indicator, especially in small helium volumes. By using a high-$T_c$ superconductor, it is possible to study the boiling of cryogenic liquids other than helium (nitrogen, oxygen, neon, hydrogen). The detector small size and thin-film design enable creating an array of sensors to study single- and multiple-bubble boiling on surfaces, also of a complex shape, optimized for improved heat removal.

## Acknowledgments

The work was supported in part by grant of the National Academy of Sciences of Ukraine UA0122U001503, SPS Programme NATO grant number G5796, grant of Slovak Republic government (application number 09I03-03-V01-00031).